%% file: main.tex
\documentclass[sigconf]{acmart}

\usepackage{tcolorbox} % import tcolorbox package
\usepackage{float}
\usepackage{textgreek}
\usepackage{listings} % for the lstlisting environment
\usepackage{xcolor} % for customizing the colors
\usepackage{subcaption}
\usepackage{tikz}
\usetikzlibrary{decorations.pathreplacing,calc,shapes,positioning,tikzmark, trees}

\definecolor{codegreen}{RGB}{70,190,120}
\definecolor{codegray}{RGB}{120,120,120}
\definecolor{codepurple}{RGB}{170,0,255}
\definecolor{backcolour}{RGB}{20,20,20}
\definecolor{codeblue}{RGB}{0,200,255}

\lstdefinestyle{mystyle}{
    backgroundcolor=\color{backcolour!80},
    basicstyle=\color{white}\ttfamily\small,
    commentstyle=\color{codegreen},
    keywordstyle=\color{codeblue},
    numberstyle=\tiny\color{codegray},
    stringstyle=\color{codepurple},
    breakatwhitespace=false,
    breaklines=true,
    captionpos=b,
    keepspaces=true,
    numbers=left,
    numbersep=5pt,
    showspaces=false,
    showstringspaces=false,
    showtabs=false,
    tabsize=4
}

\AtBeginDocument{%
  \providecommand\BibTeX{{%
    \normalfont B\kern-0.5em{\scshape i\kern-0.25em b}\kern-0.8em\TeX}}}

\setcopyright{acmcopyright}
\copyrightyear{2018}
\acmYear{2018}
\acmDOI{XXXXXXX.XXXXXXX}

\acmConference[Conference acronym 'XX]{Make sure to enter the correct
  conference title from your rights confirmation emai}{June 03--05,
  2018}{Woodstock, NY}
  
\acmBooktitle{Woodstock '18: ACM Symposium on Neural Gaze Detection,
 June 03--05, 2018, Woodstock, NY} 
\acmPrice{15.00}
\acmISBN{978-1-4503-XXXX-X/18/06}

\begin{document}

\title{Code Generation Based Grading: Evaluating an Auto-grading Mechanism for ``Explain-in-Plain-English'' Questions}

\author{David H. Smith IV}
\affiliation{%
  \institution{University of Illinois}
  \city{Urbana, IL}
  \country{USA}
}
\email{dhsmith2@illinois.edu}

\author{Craig Zilles}
\affiliation{%
  \institution{University of Illinois}
  \city{Urbana, IL}
  \country{USA}
}
\email{zilles@illinois.edu}

\renewcommand{\shortauthors}{Smith IV, D. H. and Zilles, C.}

\begin{abstract}
  
  Comprehending and elucidating the purpose of code is often cited as being a key learning objective within introductory programming courses. To address this objective ``Explain-in-Plain-English'' questions, in which students are shown a segment of code and asked to provide an abstract description of the code's purpose, have been adopted. However, given EiPE questions require a natural language response, they often require manual grading which is time-consuming for course staff and delays feedback for students. With the advent of large language models (LLMs) capable of generating code, responses to EiPE questions can be used to generate code segments, the correctness of which can then be easily verified using test cases. We refer to this approach as ``Code Generation Based Grading'' (CGBG) and in this paper we explore its agreement with human graders using EiPE responses from past exams in an introductory programming course taught in Python. Overall, we find that CGBG achieves moderate agreement with human graders with the primary area of disagreement being its leniency with respect to low-level and line-by-line descriptions of code.

\end{abstract}




\maketitle

\section{Introduction}
\input{./sections/intro.tex}

\section{Related Work}\label{sec:lit}

\subsection{``Explain in Plain English'' Questions}\label{}
\input{./sections/literature/lit-1.tex}

\subsection{Large Language Models and Introductory CS Education}\label{}
\input{./sections/literature/lit-2.tex}

\section{Code Generation Based Grading}
\input{./sections/methods/method-1.tex}

\section{Methods}\label{}
\input{./sections/methods/method-2.tex}

\section{Results}\label{}
\input{./sections/results-1.tex}

\section{Discussion}

\input{./sections/discussion.tex}

\section{Limitations}

The primary limitation of this work is the data set of responses from students
comes from a course where students are taught to respond to EiPE questions in
accordance with the rubric specified by the course staff. Students are
routinely given these questions on exams, quizzes, and homework and are thus
very familiar with the grading standards surrounding these questions.  This in
turn makes it less likely that our analysis would be able to pick up cases
where the auto-grader is grading undesirable responses as correct or vice
versa. 

\section{Conclusion}

Overall, our analysis indicates that there is a reasonable degree of agreement
between trained human graders and CGBG. It appears that much of the
disagreement comes from CGBG being more lenient than human graders,
particularly when asking students to describe code that only consists of a few
lines or operations. In instances of questions having a high quantity of false
negatives, it appears that GPT-4 struggles when students reference variables
from the original in their response which creates some ambiguity between
variable names and literal values (e.g., variable \texttt{e} vs literal
\texttt{``e''}). CGBG appears to be most effective for questions that are
complex enough they cannot tersely be described at a low level but contain
recognizable patterns (e.g., filtering, summing). With that said, as much of
the disagreement comes from GPT-4 being lenient with descriptions that are
correct but somewhat low level some instructors with less strict rubrics may
find this to be a perfectly reasonable method of grading. For carefully
designed questions and test cases, is unlikely to be more strict than a human
grader and comes with the additional layer of authenticity as it is teaching an
application of the EiPE skill which students are likely to continue to use,
code generation via LLMs such as GPT-4.

\bibliographystyle{ACM-Reference-Format}
\bibliography{refs}

\end{document}

%% file: sections/intro.tex
With the advent of large language models and the popularity they have
gained amongst the general public, many educators have raised questions and concerns
regarding the influence they may have on the future of
educational practice~\cite{pickell2023five, manoharan2023contract, xiao2022new,
cotton2023chatting, malinka2023educational}.  These concerns have been raised
due to the ease with which everything from essays to code can be generated,
which in turn draws into question the integrity of the work students
submit~\cite{cotton2023chatting}.  Despite these concerns there is wide and
growing excitement, particularly within the computer science education
community, for the potential renaissance these tools may bring by enabling new
instructional approaches~\cite{denny2022conversing, prather2023s,
becker2023programming, finnie2023my}.

Regardless of one's beliefs surrounding Large Language Models (LLMs), these tools,
particularly those models capable of code generation, are undergoing
rapid adoption to increase programming efficiency.  GitHub Copilot
has been avaliable since 2020 as a plugin for Visual Studio Code which is distributed free
to students and educators.~\footnote{www.copilot.com} Additionally, the release
of ChatGPT saw the fastest adoption rate of any platform released reaching over
a million users in just five
days.~\footnote{https://www.statista.com/chart/29174/time-to-one-million-users/}
It now seems inevitable that the usage of these tools will become as commonplace
for the next generation of computer science students as StackOverflow has been
for the last.

Given the ubiquity and ever growing presence of these tools, this draws into
question which skills students should develop in order to utilize them
proficiently. \citet{finnie2023my} have indicated that the ability to formulate
a prompt that elicits the correct response from LLMs is a skill which
may require explicit instruction. The way in which students
interact with a version of OpenAI's GPT, be it through Copilot, ChatGPT, or some other 
service, is by describing the problem statement that fits the solution they are seeking.
Thus, the problem of how to teach and evaluate students' formation of
successful queries bears some resemblance to ``Explain-in-plain-English'' (EiPE)
problems, where students are asked to describe a segment of code at a high
level.  

Existing EiPE autograders perform similarly to trained teaching assistants but
lack transparency in their grading mechanism and require human data labeling 
new question~\cite{fowler2021autograding}. To address these
limitations, we propose using LLM code generation in an autograding pipeline. This pipeline
generates code from student responses and evaluates its correctness using unit
tests in a process which we term "Code Generation Based Grading" (CGBG). CGBG not only provides
feedback by displaying the generated code and the results of unit tests, but
also streamlines the EiPE question authoring process by eliminating the need for labeling data and
training. To assess CGBG's effectiveness as an EiPE grading approach, we
investigate the following research questions:
\begin{enumerate}
  \item[\textbf{RQ1}] What is the agreement between trained human raters and code generation based grading? 
  \item[\textbf{RQ2}] What relationships exist between the features of a given question and the agreement on that question?
\end{enumerate}

%% file: sections/literature/lit-1.tex
\begin{figure}[t]
  \centering
  \includegraphics[width=\columnwidth]{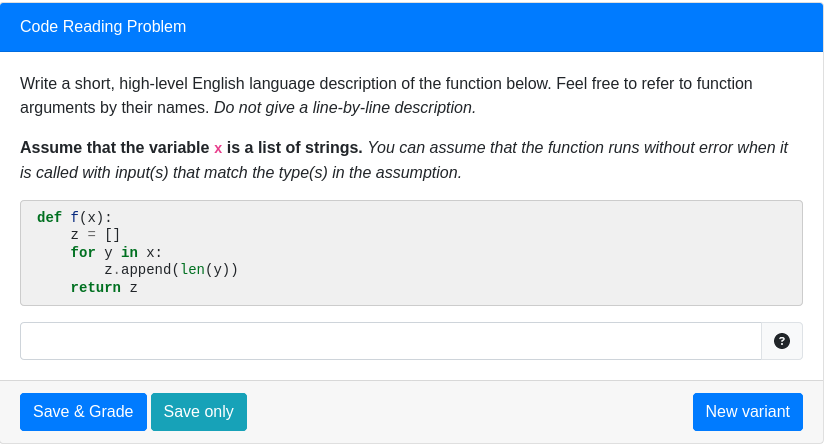}
  \caption{An example of an EiPE question as used in the course from which historical student responses were collected.}
  \label{fig:trad_eipe}
\end{figure}

The goal of EiPE questions is to evaluate
students' ability to understand and communicate the purpose of a given segment
of code~\cite{whalley2006australasian, azad2020strategies}.  In these
questions, students are presented with a segment of existing code and asked to
generate a high-level description of what the code does.  
Prior work has found performance on code comprehension tasks to be highly 
correlated with other programming skills such as code writing and 
tracing~\cite{lister2009further, venables2009closer, fowler2022reevaluating,lopez2008relationships}. 
\citet{xie2019theory} has suggested a sequenced approach wherein students are
introduced to programming concepts incrementally the ability to describe code
is the penultimate step before code writing. In this way, gaining proficiency
in articulating the purpose of code may be placed amongst the pantheon
of programming skills introductory courses seek to impart on their students.

However, despite the popularity of EiPE questions, they have been historically
difficult to scale to larger classrooms as they typically require manual
grading and the development of complicated and subjective
rubrics which can be difficult to apply to student
responses~\cite{azad2020strategies}.  The issue of defining an EiPE rubric has
been addressed by multiple studies, each attempting to evaluate the ability of
students to form an unambiguous, functionally correct, and abstract description
of code~\cite{chen2020validated}. In a study investigating faculty's
perceptions on grading standards for EiPE questions, \citet{fowler2021should}
found that faculty acknowledged the imprecision of natural language
descriptions and placed more value on correctness than instances of slight
ambiguity and, in general, preferred high-level descriptions over low-level
ones.

Despite the strides that have been made in defining a rubric for EiPE
questions, the issue of scale remains. Prior work by
\citet{fowler2021autograding} addressed this issue by developing automated
grading systems for EiPE questions. Those systems produce results similar
to that a trained human grader. However, this system does suffer in a formative
context as it is unable to provide feedback on why a student's response
was graded as correct or incorrect. Additionally, it requires a large corpus of
human labeled training data that must be constructed on a per question basis
which adds a layer of difficulty to the process of creating EiPE questions.
Given these two shortcomings, this leaves open the door for systems that seek
to add both a layer of transparency to the grading process and streamline the
process of creating auto-grading mechanisms for EiPE questions.

%% file: sections/literature/lit-2.tex
Large language models (LLMs) have already been demonstrated to be a powerful
and effective tool for generating code in a variety of contexts, often
performing at or above that of the average student on introductory programming
problems~\cite{denny2022conversing}.  A study by \citet{finnie2022robots}
evaluated the responses Copilot generated to introductory CS problem, including
the (in)famous rainfall problem~\cite{wermelinger2023using}.  The system proved
extremely successful, generating a wide variety of solutions to each problem
type. Even beyond standard programming problems, LLMs have been shown to be
successful, albeit to a lesser extent, at solving Parsons
problems~\cite{reeves2023evaluating}.  Though many of these studies make note
of the existential threat this provides to the current state of CS education
(introductory CS education in particular) they also make note of the avenues
that are likely to emerge in the future for leveraging such technologies for
teaching.

In a recent study \citet{denny2022conversing} evaluated how various approaches
to prompt engineering for CS1 questions impacted GitHub Co-Pilots correctness.
In doing so, they found that Copilot successfully generated solutions for approximately 80\%
of the problems within two attempts. Notably, they found that the two
categories of prompts that Copilot had the most difficulty with were those that
were both too abstract and too verbose. Within the context of the ``Structure
of Observed Learning Outcomes'' (SOLO) taxonomy and past work evaluating EiPE
questions, this may suggest that terse, multi-structural responses are the most
effective at eliciting correct responses from Copilot~\cite{lister2006not}.  Furthermore,
\citet{chen2021evaluating} found that the correctness of Copilot's 
responses decreased as the length of the prompts increased.  This suggests there is an alignment between the successful prompts to LLMs
and the canonically ``correct'' EiPE responses.

%% file: sections/methods/method-1.tex
\begin{figure*}
  \input{./imgs/pipeline.tex}
  \caption{The process of generating code from a prompt and grading it.}
  \label{fig:pipeline}
\end{figure*}
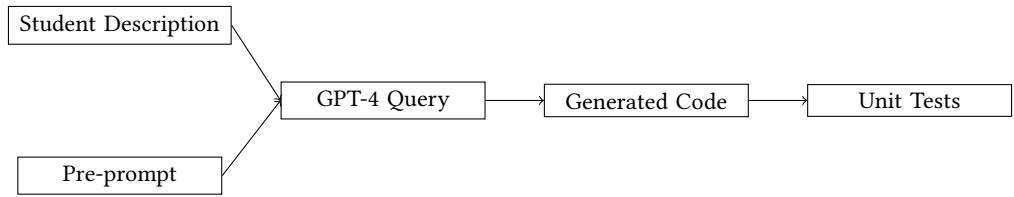

The ``Code Generation Based Grading'' (CGBG) process we propose (Figure~\ref{fig:pipeline}) is divided into three distinct
steps. First, a student's response to an EiPE question and a pre-prompt are combined
and fed to GPT-4 to generate a function. That function is then run against a set of 
unit tests which are manually defined for each question. The students' grades are dependent on the unit tests, receiving a 1 if all tests pass and 0 if any fail. The
pre-prompt fromat used in this study is as follows: 
\begingroup
\addtolength\leftmargini{-0.2in}
\begin{quote}
  Pretend you are an introductory programming student with a rudimentary understanding
  of programming. You are proficient in the use of functions, loops, conditionals, basic
  builtin data structures (i.e., list, sets, dictionaries), and user input/output. Construct
  code as response to the following prompt: Generate a function ``studentCode''
  that \mbox{\textit{<StudentResponse>}}.  Respond with the code only and no other explanatory
  text or example test cases.
\end{quote}
\endgroup

Here, the pre-prompt instructs GPT-4 to pretend to be an introductory student with the intention
of it generating code would be interpretable by students and, thus,
could serve as feedback to the students with each submission.  
The pre-prompt also includes instructions to create a function named ``studentCode''
to ensure a standardized function name for testing. The student's EiPE response is
 used to replace \texttt{<StudentResponse>}. 

This pipeline simplifies the question authoring process for automatically graded
EiPE quesitons to that of traditional code writing questions. Previously developed EiPE autograders require the creation of human labeled
datasets and training of NLP classification models~\cite{azad2020lessons}.  This approach also offers
the benefit of allowing the instructor to concretely define what aspects of the
code and edge cases should be described in the prompt through the
unit tests. For example, consider the following code:
\begin{lstlisting}[language=Python]
def foo(x, y):
  for i in x:
    if i == y:
      return y
  return -1
\end{lstlisting}
If the instructor's goal is to ensure that an
explanation for the following code includes an exact description of the function's return behaviour that can be verified via a test case.

With this process defined there is one final consideration: how best to 
accommodate the typically non-deterministic nature of GPT-4. 
Non-deterministic, in this context, means that given the same prompt GPT-4 may
generate a different response each time it is queried. Through the model's temperature parameter, the
user can control the ``creativity'' of the model, with a temperature of 0.0
being deterministic and 1.0 being the most creative. What we wish to avoid is a student 
submitting a response which would generally generate correct code being penalized 
due to GPT-4 generating an overly creative interpretation of their prompt. 
To evaluate the best way to prevent such erroneous grading, we put forth and compare 
the following three methods of grading:
\begin{itemize}
  \item \textbf{Correct at 0.0 Temperature:} Temperature of GPT-4 is set to 0.0 such that the response for a given prompt is deterministic. This single response is then used for grading.
  \item \textbf{Best of 5 at Temperature 0.5:} For a given prompt, GPT-4 is queried 5 times at a temperature of 0.5. If at least one of these responses passes all test cases the student is awarded full marks; otherwise they receive a 0.
  \item \textbf{Majority Vote at Temperature 0.5:} GPT-4 is queried 5 times at a temperature of 0.5. If at least 3 of the 5 responses pass all test cases students are awarded full marks; otherwise, they receive a 0.
\end{itemize}

\begin{figure*}
  \subfloat[Single Response at Temp=0]{
    \includegraphics[width=0.3\textwidth]{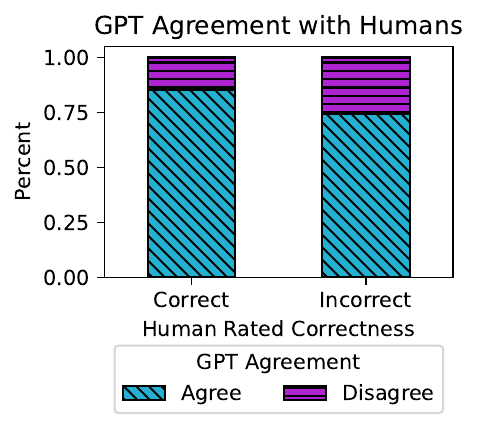}
    \label{fig:0agree}}
  \hfill
  \subfloat[Majority Vote of 5 Responses at Temp=0.5]{
    \includegraphics[width=0.3\textwidth]{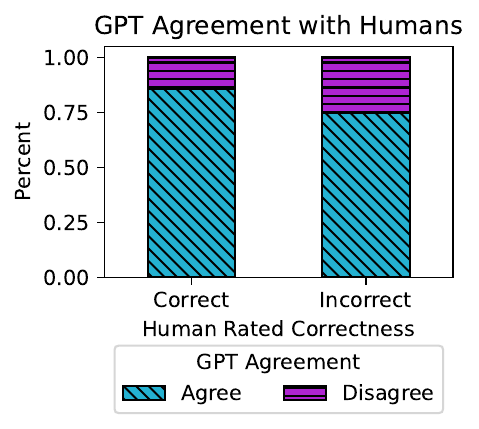}
    \label{fig:mvote}}
  \hfill
  \subfloat[Best of 5 Responses at Temp=0.5]{
    \includegraphics[width=0.3\textwidth]{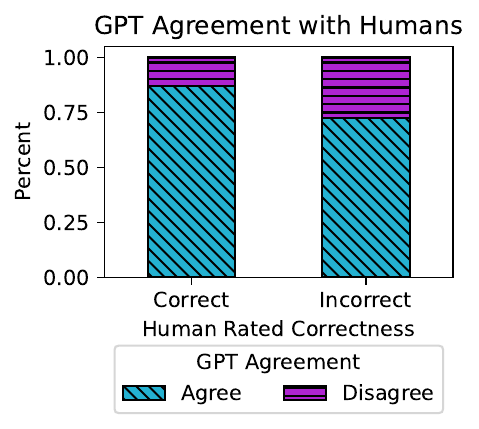}
    \label{fig:avote}}
  \caption{The agreement between the three approaches to grading with CGBG and human grading. Overall, all appear to have nearly identical levels of false positives and false negatives.}
  \label{fig:agree}
\end{figure*}

%% file: imgs/pipeline.tex
\begin{tikzpicture}[node distance=2cm]

    \node (sd) [draw, text width=2.75cm, align=center] {Student Description};
    \node (preprompt) [draw,  below of=sd, text width=2.5cm, align=center] {Pre-prompt};

    \node (query) [draw,  right of=preprompt, yshift=1cm, xshift=1.5cm, text width=2.5cm, align=center]{GPT-4 Query};

    \node (generate) [draw,  right of=query, xshift=1.5cm, text width=2.5cm, align=center]{Generated Code};
    \node (test) [draw,  right of=generate, xshift=1.5cm, text width=2.5cm, align=center]{Unit Tests};

    %% Arrows
    \draw[->] (sd.east) -- (query.west);
    \draw[->] (preprompt.east) -- (query.west);
    \draw[->] (query.east) -- (generate.west);
    \draw[->] (generate.east) -- (test.west);

\end{tikzpicture}

%% file: sections/methods/method-2.tex
\begin{table}[t]
  \centering
  \begin{tabular}{|c|c|}
    \hline
    Kappa Value & Agreement \\ \hline
    $<0$ & Less than chance agreement \\ \hline
    $0.01-0.20$ & Slight agreement \\ \hline
    $0.21-0.40$ & Fair agreement \\ \hline
    $0.41-0.60$ & Moderate agreement \\ \hline
    $0.61-0.80$ & Substantial agreement \\ \hline
    $0.81-0.99$ & Almost perfect agreement \\ \hline
    $1.00$ & Perfect agreement \\ \hline
  \end{tabular}
  \caption{Standard Cutoffs for Interpreting Cohen's $\kappa$}
  \label{tab:kappa}
  \vspace{-.75cm}
\end{table}

To determine the agreement between human raters and the test cases run on the
code generated by GPT-4, we begin with data collected from exam data in an
introductory programming course for non-technical majors at a large research 
university in the United States. EiPE questions have been used routinely in the
course's proctored exams and have historically been manually graded by the
course staff. Typically, TAs for the course are trained on grading EiPE responses 
by longstanding members of the course staff with significant experience with grading  
EiPE questions. This training process involves grading some answers and meeting 
to discuss the outcomes. The EiPE responses were typically graded by at 
least two TAs, with disagreements reconciled to decrease the likelihood of incorrect 
grading results. The course staff grades these question in accordance with a rubric
informed by the literature covered in Section~\ref{sec:lit}.  This rubric is
composed of the following three criteria:
\begin{itemize}
  \item \textbf{Correctness:} The student must describe the process fully such
    that the process they are describing is functionally correct. If a student
    is describing a function that filters odd numbers and they incorrectly
    describe it as filtering even numbers, the response is graded as incorrect.
  \item \textbf{Unabmiguous:} A student must describe the process fully such
    that multiple, contradictory interpretations of the code is not possible.
    For example, if a student describes a function as filtering numbers, but
    does not describe the criteria for filtering, this description would
    be considered ambiguous.
  \item \textbf{High-Level:} The goal for these EiPE questions is for students
    to describe the general purpose of code snippets at a high level of
    abstraction, not provide a line-by-line description.  For example, when
    describing a function that determines if a number is prime, a student
    should simply state that the function determines if a number is prime,
    rather than describing the process of iterating over every number less than
    the input and checking if it is divisible by the input.
\end{itemize}
These questions are graded on a binary scale where a correct answer receives
full marks for meeting all three of these criteria. In total, 6380 student
responses from 42 EiPE questions were included in the analysis.

We use Cohen's \textkappa{} to measure the agreement between the human raters and each
of the proposed GPT grading approaches.  Cohen's \textkappa{} is a measure of commonly used
to evaluate interrater reliability.  Interpreting
the \textkappa{} statistic is traditionally done using the scale shown in 
Table~\ref{tab:kappa}~\cite{mchugh2012interrater}.  In our case, we use the same scale to interpret the
agreement between the human graders and the results of the test cases run on
the code generated by GPT-4.

%% file: sections/results-1.tex
\begin{figure}[t]
  \centering
  \includegraphics[width=200px]{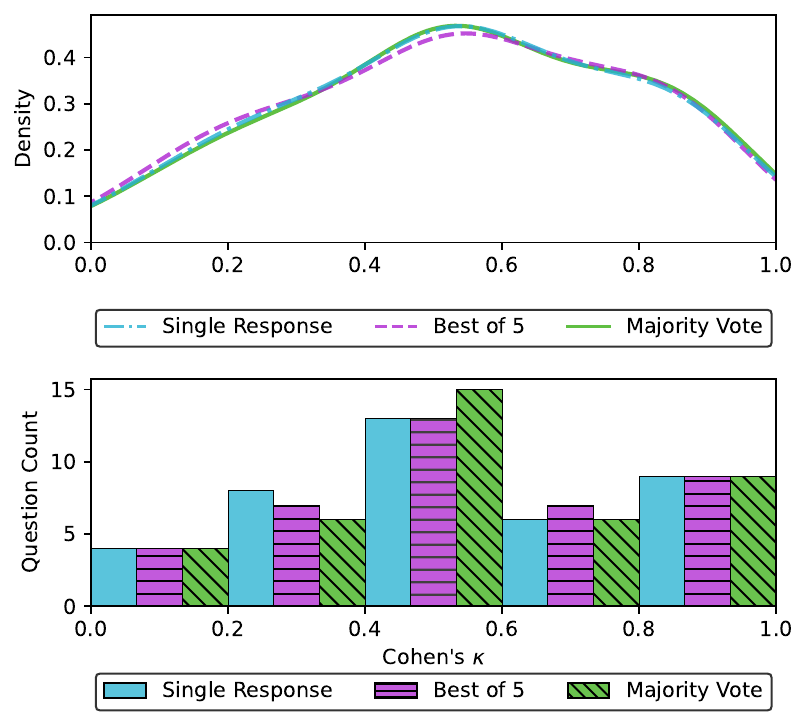}
  \caption{Cohen's $\kappa$ was computed for each question within each of the three grading methods. Results appear similar for each method with the majority achieving moderate agreement.}
  \label{fig:kappa}
  %\vspace{-0.25cm}
\end{figure}

The overall agreement between human graders and CGBG across all questions was similar 
for each of the three CGBG methods. Grading a single response from GPT-4 at temperature  
0 and grading based on majority vote of five responses generated at a temperature of 0.5 
both achieved $\kappa=.58$.  Grading based on the best of five responses at a temperature of 0.5
was slightly lower at $\kappa=.57$. When looking at where this disagreement occurs for each of the
three CGBG methods it appears the majority occurs when CGBG grades a response as correct when 
human graders graded it as incorrect (Figure~\ref{fig:agree}). This indicates that CGBG has a
reasonable agreement with human graders, but, when disagreement occurs, it is
more likely to be lenient than strict. 

Agreement between human graders and each of the three CGBG methods was computed for 
each of the questions. The majority of the questions achieved a moderate agreement
and all CGBG methods achieved similar results (Figure~\ref{fig:kappa}). 
of the grading method used (Figure~\ref{fig:agree}). 
To gain a better understanding of where CGBG fails and succeeds
we will next look at questions that achieve low, moderate, and high agreement
with human graders. Given the similarity of the results between the three grading methods, 
we will use the results of ``correct at 0.0 temperature'' for this portion of the 
analysis.

\subsection{Questions with Low Agreement ($\kappa \leq 0.4$)}\hfill

\paragraph{``Simple'' Questions: } Most notable among those questions with low agreement are those where the code 
students were asked to describe was simple. Such questions occur early in the course prior to the introduction
of loops and more complex conditional structures. As such, these questions typically involve a function definition
and a single line of code which can be described tersely without illustrating the ``high-level'' purpose of the code.
For example, one question asked students to describe the following function which simply  returns the average 
of a list of numbers.

% Give me a pretty lstlisting for this
\begin{lstlisting}[language=Python]
def foo(lst):
  return sum(lst)/len(lst)
\end{lstlisting}

\begin{figure}[t]
  \centering
  \includegraphics[width=\columnwidth]{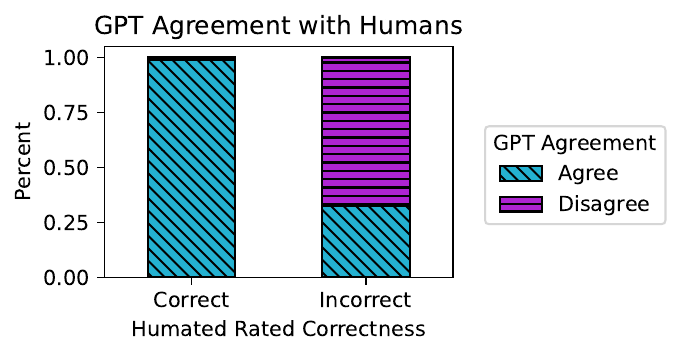}
  \caption{Agreement between human graders and CGBG on questions with a simple structure. Overall, there appears to be a very low false negative rate but a high false positive rate. This is likely due to students being able to provide functionally correct but low level responses to these questions.}
  \label{fig:simpleqidagree}
\end{figure}

In accordance with the rubric used in the course, descriptions of the above code
would be required to be high-level descriptions in order to be marked incorrect.
For example, a student response that simply states the function
returns ``\textit{the sum of the element in the list divided by the length of
the list}'' would be marked as incorrect by human graders as it does not 
describe the function as ``\textit{finding the average}''. However, GPT-4 will 
generate functionally correct code and thus would mark this response as correct.
Other questions that suffered from this issue included:
\begin{itemize}
  \item An absolute value function.
  \item Returning the maximum of two numbers.
  \item Determining if x is a multiple of y.
  \item Returning the range (difference between max and min) of a list of numbers.
  \item Returning the maximum of two numbers.
\end{itemize}
Overall, the code generation based grading method was much more lenient
than human graders on the aforementioned questions (Figure~\ref{fig:simpleqidagree}).
Looking at the agreement between human graders and GPT on these questions in
the takeaways from this analysis will differ depending on the goal of the rubric
being used.  If an instructor finds functionally correct but high-level
responses to these early questions to be acceptable, then the issue of
this grading method being too lenient may not be considered a problem. However, if applying
a strict requirement that even simple questions must be answered in a high-level
manner, then it would be insufficient to rely on CGBG alone in such cases. Other models
or heuristics may need to be placed along CGBG to evaluate difference facets of ``correctness''. 

\subsection{Questions with Moderate Agreement ($ 0.6 \geq \kappa > 0.40$)}\hfill

% make a figure
\begin{figure}[t]
  \includegraphics[width=\columnwidth]{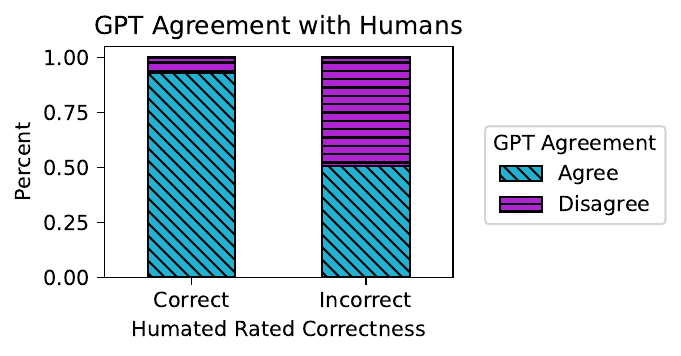}
  \caption{Questions with simple \textit{but recognizable} structure that achieved moderate agreement
  between CGBG and human graders.  We see a similar trend to previous simple
  questions in that there is a high false positive rate that likely results from
  students describing this code at a low level.}
  \label{fig:mediumsimple}
\end{figure}

% Simple but recgonizable patterns
\paragraph{Simple but Recognizable Patterns:} Similar to the prior section,
questions with moderate agreement also struggled with responses that were
functionally correct but not sufficiently high-level being graded as correct by
CGBG (Figure~\ref{fig:mediumsimple}). For example, responses describing the following code
\begin{lstlisting}[language=Python]
def foo(x):
  return x % 2 == 0
\end{lstlisting}
included:
\begin{itemize}
  \item ``\textit{Returns true if the remainder of dividing 2 from x equals 1}''
  \item ``\textit{True if modulo of x by 2 is 1 false otherwise}''
\end{itemize}
The human graders mark responses such as these incorrect as they do not
describe the higher level purpose of determining if a number is even. Other
questions that suffered from this issue included a function that determines if
a number is odd and another that determines if a list is empty.

What distinguishes these questions from those covered in the former section is
these code snippets cover patterns that students routinely encountered in the
course. As such, the superior \textkappa{} does may not be driven by the GPT model being more likely to
distinguish between high and low level responses. Rather, these patterns are 
so common that students are likely to recognize them and use high-level language
in their responses, reducing the overall number of low level responses for these questions.

\begin{figure}[t]
  \includegraphics[width=\columnwidth]{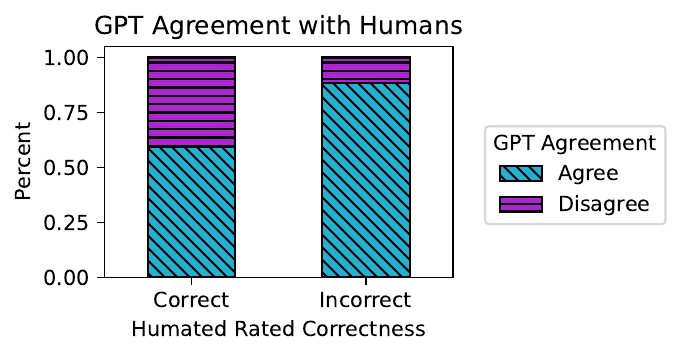}
  \caption{Agreement for questions with moderate agreement between CGBG and human ground truth that suffer from high false negative rates.}
  \label{fig:highfn}
\end{figure}

\paragraph{Context Issues:} 
The only category of questions that had a high false negative rate were those
wherein students refer to variables from the code snippet in their
responses on questions involving string manipulation
(Figure~\ref{fig:highfn}).  Specifically, a small number of
questions that involved slicing ordered collections or iterating between ranges commonly ran into issues with referring to variables by name. For example, consider the following function:
\begin{lstlisting}[language=Python]
def foo(x, y, z):
  return x[x.index(y)+1: x.index(z)]
\end{lstlisting}
A response of the form ``\textit{return the substring that is in between y and
z }'' would be marked as correct by human graders. Given that GPT-4 is generating code 
based solely on the student's prompt there exists some ambiguity in what y and z 
refer to. As such, GPT often generates code that assumes y and z are letters in the
string rather than variables and slices between them. For example, GPT-4 
generated the following code from such a response,
\begin{lstlisting}[language=Python]
def foo(str):
  return str[str.index("y")+1:str.index("z")]
\end{lstlisting}
It appears that this issue is specific to questions involving strings as when 
students refer to variables in problem involving math (e.g., check if x is even)
GPT-4 correctly determines that x must be a variable containing a number. With that said,
this issue could be mitigated by either providing more context to the GPT-4 model in the 
pre-prompt or instructing students not to refer to variables in their responses.

\subsection{Questions with High Agreement ($\kappa > 0.6$)}

It appears those questions with high agreement lack 
clusters of traits that can be used to explain the agreement. Instead,
there appears to be a general trend of questions that have sufficiently
complex code that it can not be explained in a line-by-line fashion. For
example, the following questions all had \textkappa{} values above 0.8:

\begin{lstlisting}[language=Python]
def foo(x, y):
    for z in x:
        if z == y:
            return True
    return False

\end{lstlisting}
\begin{lstlisting}[language=Python]
def foo(x):
    total = 0
    for i in x:
        if i > 0:
            total += i
    return total
\end{lstlisting}

In addition to being sufficiently complex, problems which achieved high
agreement generally followed programming patterns that students are explicitly
taught in the course. As such, students may be more likely to recognize each of
these problems as belonging to those patterns and use correct and sufficiently
high-level language to describe them. This would satisfy the rubric used by
human graders and be sufficient to generate the correct code using GPT-4.

%% file: sections/discussion.tex
Overall, the CGBG approaches introduced in this paper all achieve a moderate agreement
with human raters. The primarily limitation appears to be that this method of 
grading is unable to distinguish between high and low level descriptions and 
is thus more lenient, particularly on responses describing shorter segments of 
code. As such, deployment of this method of grading might be most appropriate for 
longer and more complex segments of code if ensuring students answer with a high-level
description is a concern. Additional heuristics such as limiting the length of a 
students response may also be used to help ensure this grading standard is met. 

However, looking beyond just the accuracy of the grader, there are several
other affordances that should be considered. The first is the ease with which
new questions can be authored. Using this method of grading, writing an EiPE
question becomes a simple as writing test cases and a segment of sample code.
This also allows the question author to control the level of detail students should
provide in their response. For example, if the handling of an edge case should
be explicitly mentioned, a test case can be written to ensure it is accounted
for.

Auto-graded EiPE questions have also historically suffered from a lack of
transparency in their grading mechanism. This can lead to students distrusting
the accuracy of the grades provided by the system and thus hinder their ability
to improve their ability in answering said questions~\cite{hsu2021attitudes}.
Through this grading mechanism, students can be given the code that was
generated as feedback to give insight into how GPT-4 is interpreting their
description and test cases to highlight specific inputs that failed to produce
a desired output. Future work should explore the impact of this feedback on
students' ability to improve their performance on these questions as well as
students perceptions surrounding the grader's accuracy.